\addcontentsline{toc}{section}{References}

\documentclass[runningheads]{svmult}

\usepackage{makeidx}   % allows index generation
\usepackage{graphicx}  % standard LaTeX graphics tool
\usepackage{subeqnar}  % subnumbers individual equations
\usepackage{multicol}  % used for the two-column index
\usepackage{physprbb}  % modified textarea for proceedings,
\makeindex             % used for the subject index
\begin{document}
\title*{Surface features, rotation and atmospheric variability of ultra cool dwarfs}
%
%
%\toctitle{Surface features, rotation and atmospheric variability of ultra cool dwarfs}
% allows explicit linebreak for the table of content
%
%
%\titlerunning{Surface features, rotation and atmospheric variability of ultra cool dwarfs}
% allows abbreviation of title, if the full title is too long
% to fit in the running head
%
\author{C.A.L. Bailer-Jones}
\authorrunning{C.A.L. Bailer-Jones}
% if there are more than two authors,
% please abbreviate author list for running head
%
%
\institute{Max-Planck-Institut f\"ur Astronomie, K\"onigstuhl 17, D-69117 Heidelberg, Germany\\email: calj@mpia-hd.mpg.de}

\maketitle              % typesets the title of the contribution

\begin{abstract}
Photometric $I$ band light curves of 21 ultra cool M and L dwarfs are
presented. Variability\index{variability} with amplitudes of 0.01 to
0.055 magnitudes (RMS) with typical timescales of an hour to several
hours are discovered in half of these objects.  Periodic variability
is discovered in a few cases, but interestingly several variable
objects show no significant periods, even though the observations were
almost certainly sensitive to the expected rotation periods.  It is
argued that in these cases the variability is due to the evolution of
the surface features on timescales of a few hours. This is supported
in the case of 2M1145 for which no common period is found in two
separate light curves.  It is speculated that these features are
photospheric dust clouds, with their evolution possibly driven by
rotation and turbulence. An alternative possibility is
magnetically-induced surface features.  However, chromospheric
activity undergoes a sharp decrease between M7 and L1, whereas
a greater occurrence of variability is observed in objects later than
M9, lending support to the dust interpretation.

\vspace*{1ex}
\end{abstract}

\begin{minipage}{10.7cm}
\noindent
To appear in {\it Ultracool Dwarf Stars} (Lecture Notes in Physics), H.R.A.\ Jones, I.\ Steele (eds), Springer-Verlag, 2001
\end{minipage}

\section{Introduction}

Large numbers of very cool compact objects -- namely low mass stars,
brown dwarfs and giant gas planets -- have only recently become
available through large scale surveys, particularly in the far red and
near infrared where these objects radiate most of their energy.  This
has presented the opportunity to systematically study their intrinsic
properties. Indeed, the L dwarf sequence has recently been introduced
to account for the increasing number of objects found with effective
temperatures apparently in the range 2200--1300\,K
\cite{basrietal00}\cite{kirk99}\cite{kirkpatrick00}\cite{martin99},
and the T dwarf class covers the even cooler objects (similar to
Gl\,229B) now being discovered
\cite{burgasser00}\cite{leggett00}\cite{noll00}.

For a number of reasons, these ultra cool dwarfs are likely to have
interesting and complex atmospheres\index{atmospheres}.  First, they are fully
convective.  Second, many are rapid rotators
\cite{basrietal00}. Third, at these low temperatures, solid dust
particles, or condensates, form. These first two properties may be a
driver for atmospheric dynamics, or weather, and the presence of dust
makes it possible that large-scale clouds form. Although these
objects cannot (yet) be resolved, the presence of weather patterns can
be investigated by via accurate photometric monitoring, which is the subject
of this article.

\subsection{Previous Work}

Initial attempts to observe variability in ultra cool dwarfs have met with mixed
results. Terndrup et al.\ \cite{terndrup99} searched for rotational
modulation of the light curves of eight M-type stars and brown dwarfs
in the Pleiades.  They derived periodicities for two low mass stars,
but found no significant variability in the rest of the sample. Tinney
\& Tolley \cite{tinney99} found some evidence for variability in an M9
brown dwarf with an amplitude of 0.04 magnitudes over a few hours, but
detected no variability above 0.1 magnitudes in an L5 dwarf.  Nakajima
et al.\ \cite{nakajima00} found variability in the near infrared
spectrum of a T dwarf over a period of 80 minutes.  In a precursor to
the present project, Bailer-Jones \& Mundt \cite{bailer-jones99}
observed six M and L dwarfs and found variability in one (2M1145), to
which a tentative period of 7.1 hours was assigned (pending
confirmation).

\subsection{Observational Sample}

\index{M dwarf}\index{L dwarf}\index{brown dwarf}\index{star, very low mass}
This paper reports results from an observational program to monitor
photometric $I$ band variability in a sample of 21 late M and L dwarfs
(Table~\ref{targets}).  Only about 30 L dwarfs were known at
the time of the observations, thus greatly limiting the choice of
targets. Objects brighter than $I=19.0$ were
preferentially selected, but there are no other (known) selection
biases. Ten of the targets are field dwarfs discovered by the Two
Micron All Sky Survey (2MASS) and the Sloan Digital Sky Survey
(SDSS). Five objects are members of the Pleiades (age 120 Myr): Teide
1 and Calar 3 are confirmed brown dwarfs, Roque 11 and 12 are probably
brown dwarfs, and Roque 16 is very close to the hydrogen burning limit
(so its status is uncertain).  The six remaining objects are candidate
members of the $\sigma$ Orionis cluster (age 1--5 Myr), with masses
between 0.02 and 0.12\,$M_{\odot}$.

\begin{table*}
\caption[]{Properties of the ultra cool dwarf targets.  Each
reference makes use of a different $I$ band and even definition of
magnitude, so values are only intended to be indicative. In
particular, the SDSS $I$ filter is somewhat bluer than the Cousins $I$
filter, thus yielding fainter magnitudes for L dwarfs. The spectral
types in parentheses have been estimated from the $R-I$ colours in
\cite{bejar99}.
\label{targets}
}
\begin{tabular}{lllllll}
\hline
name        	& IAU name               	& $I$	& SpT 	& H$\alpha$ EW	& LiI EW	& ref	\\
            	&                        	&	&   	& \AA\   	& \AA\          & 	\\
\hline
2M0030		& 2MASSW J0030438$+$313932	& 18.82	& L2	& 4.4 $\pm$0.2	& $< 1.0$			& \cite{kirk99}	\\
2M0326		& 2MASSW J0326137$+$295015	& 19.17	& L3.5	& 9.1$\pm$0.2	& $< 1.0$			& \cite{kirk99}	\\
2M0345		& 2MASSW J0345432$+$254023	& 16.98	& L0	& $\leq 0.3$	& $< 0.5$			& \cite{kirk99}	\\
2M0913		& 2MASSW J0913032$+$184150    	& 19.07	& L3	& $< 0.8$	& $< 1.0$			& \cite{kirk99}	\\
2M1145		& 2MASSW J1145572$+$231730    	& 18.62	& L1.5	& 4.2$\pm$0.2	& $< 0.4$			& \cite{kirk99}	\\
2M1146		& 2MASSW J1146345$+$223053    	& 17.62	& L3	& $\leq 0.3$	& 5.1$\pm$0.2~			& \cite{kirk99}	\\
2M1334		& 2MASSW J1334062$+$194034	& 18.76	& L1.5	& 4.2$\pm$0.2	& $< 1.5$			& \cite{kirk99}	\\
2M1439		& 2MASSW J1439284$+$192915	& 16.02 & L1	& 1.13$\pm$0.05& $< 0.05$			& \cite{reid00}	\\
SDSS~0539~	& SDSSp  J053951.99$-$005902.0~	& 19.04~ & L5	& 		&				& \cite{fan00}	\\
SDSS~1203	& SDSSp  J120358.19$+$001550.3	& 18.88 & L3	& 		&				& \cite{fan00}	\\
Calar 3 	&               		& 18.73	& M9	& 6.5--10.2	& 1.8$\pm$0.4			& \cite{rebolo96}	\\
Roque 11	& RPL J034712$+$2428.5 		& 18.75	& M8	& 5.8$\pm$1.0	&				& \cite{zap99}	\\
Roque 12	& 				& 18.47	& M7.5	& 19.7$\pm$0.3	& $\leq 1.5$			& \cite{martin98}	\\
Roque 16	& RPL J034739$+$2436.4		& 17.79	& M6	& 5.0$\pm$1.0	& 				& \cite{zap99}	\\
Teide 1 	& TPL J034718$+$2422.5 		& 18.80	& M8	& 3.5--8.6	& 1.0$\pm$0.2			& \cite{rebolo95}	\\
S~Ori~31	& S Ori J053820.8$-$024613	& 17.31	& (M6.5)~	&	&				& \cite{bejar99} \\
S~Ori~33	& S Ori J053657.9$-$023522	& 17.38	& (M6.5)	&	&				& \cite{bejar99} \\
S~Ori~34	& S Ori J053707.1$-$023246	& 17.46	& (M6)	& $\leq 5.0$&					& \cite{bejar99} \\
S~Ori~44	& S Ori J053807.0$-$024321	& 19.39	& M6.5	& 60.0$\pm$1.0	&				& \cite{bejar99} \\
S~Ori~45	& S Ori J053825.5$-$024836	& 19.59	& M8.5	&		&				& \cite{bejar99} \\
S~Ori~46	& S Ori J053651.7$-$023254	& 19.82	& (M8.5)	&	&				& \cite{bejar99} \\
\hline
\end{tabular}
\end{table*}

\section{Observations and Data Reduction}\label{data}

The targets were observed with a CCD camera on the 2.2m telescope at
the Calar Alto Observatory (Spain) over three periods: January 1999
(AJD 1187.4--1192.8, hereafter 99-01), September 1999 (AJD
1432.8--1436.2, hereafter 99-09) and February 2000 (AJD
1601.8--1607.2, hereafter 00-02). 
AJD=JD-2450000. Exposure times of 8 minutes were used for most objects
to achieve a signal-to-noise ratio (SNR) for the target objects of at
least 100.  Objects were observed repeatedly each night for several
nights to construct a light curve with typically 30 points.  The data
reduction consisted of careful flat fielding and fringe removal to
reduce all errors to less than 0.5\%. More details are given in
Bailer-Jones \& Mundt \cite{bjm00}.

\section{Light Curve Analysis: Theory}

Differential light curves\index{light curve} were obtained for each target relative to a
number of reference stars in the field. These reference stars were
chosen to be bright and isolated.  Fluxes for all objects and
reference stars were determined using aperture photometry with an
aperture radius of 3.5 pixels ($1.9''$).  The average flux of the
reference stars forms a magnitude against which fluctuations in the
target are monitored.  The zero-mean differential light curve for each
target, consisting of $K$ points (or epochs), is denoted
$m_d(1),m_d(2),\ldots,m_d(k),\ldots,m_d(K)$. The {\em total}
photometric error at each point, $\delta m_d(k)$, has been carefully
determined considering photon-noise statistics plus contributions from
imperfect flat fielding and fringe correction.

\subsection{$\chi^2$ Test and Reference Star Rejection}\label{chisqtest}

A general test of variability is made using a $\chi^2$ test, in
which we evaluate the probability, $p$, that the deviations in the light
curve are consistent with the photometric errors. 
The statistic is
\begin{equation}
\chi^2 = \sum_k^K \left( \frac{m_d(k)}{\delta m_d(k)} \right)^2
\end{equation}
such that a large $\chi^2$ indicates greater deviation compared to the
errors, and thus a smaller $p$. An object is considered variable if
$p<0.01$ (a 99\% confidence level).

This test is first used to iteratively remove variable reference
stars, by calculating the light curve for each reference star relative
to all the other reference stars. The most variable reference star is
removed from the reference list, and the procedure repeated until only
non-variable reference stars remain (i.e.\ those with $p>0.01$).  Due
to the large number of reference stars used (typically 20--30), any
residual low-level variability in any one object will be greatly reduced in
the averaged reference level.

The reliability of the $\chi^2$ test depends on an accurate
determination of the magnitude errors in the target.  This has been
confirmed, as objects of similar brightness to the target have
variations no larger than the photometric errors for the target, thus
ensuring that the errors have not been overestimated.

\subsection{Power Spectrum Analysis}

Evidence for {\em periodic} variability was searched for using the
{\em power spectrum}\index{power spectrum} or periodogram. A dominant
periodicity may be present at the rotation period due to rotational
modulation of the light curve by surface inhomogeneities.  The power
spectrum of a continuous light curve, $g(t)$, is $|G(\nu)|^2$, where
$G(\nu) = {\rm FT}[g]$ and FT denotes a Fourier transform. However,
the target objects are only observed at the discrete time intervals
given by the sampling function, $s(t)$. Thus the power spectrum of the
measured {\it discrete} light curve, $d(t) = g(t)s(t)$, is
$|D(\nu)|^2$, where $D(\nu) = {\rm FT}[d(t)] = G(\nu) \otimes W(\nu)$.
$W(\nu)={\rm FT}[s(t)]$ is the {\it spectral window function} and
$\otimes$ is the convolution operator. Thus the measured power
spectrum, $|D(\nu)|^2$, may show spurious features due to the way in
which the true continuous light curve was sampled. Such features are
not intrinsic to the source and may obscure features which are
\cite{deeming75}\cite{roberts87}, particularly at the low SNRs
considered here.

However, it is possible to estimate $G(\nu)$ (power and phases) from
the raw or {\it dirty} power spectrum through an iterative
deconvolution using the {\sc CLEAN} algorithm, which was first introduced to
reconstruct aperture synthesis data in radio astronomy
\cite{roberts87}. The resulting {\it cleaned} spectrum generally
consists of peaks at a number of distinct frequencies, plus a residual
spectrum and noise. The {\sc CLEAN} code used calculates the power and
phases of the components in the window function and the cleaned power
spectrum \cite{lehto00}.  The noise in the power spectrum is
calculated from the photometric errors and the time sampling, and is
stated in the captions to the power spectra in the next section.
Peaks which are not more than several times this noise level should
not be considered significant.  The uncertainty in a determined
frequency in the power spectrum is set by t$_{\rm max}$, and is
approximately $\tau^2/(2 t_{\rm max})$ for a period $\tau$. However,
for very short periods the error is constant due to the finite
integration time.  The longest period to which the observations are
sensitive is of order $t_{\rm max}$.

\section{Results}\label{results}

\subsection{General Results}\label{genresults}

The results of the application of the $\chi^2$ test to the 21 targets
are shown in Table~\ref{detections} for the detections and
Table~\ref{nondetections} for the non-detections of
variability.  For those objects in which we did not detect
variability, we have set upper limits on the amplitude. This was done
by creating a set of synthetic light curves by multiplying each
$m_d(k)$ by $1+a$, for increasing (small) values of $a$. The amplitude
limits were obtained from that synthetic light curve which gave
$p=0.01$ according to the $\chi^2$ test. Note that a number
of detections are close to the confidence limit of $p=0.01$, so the
division between Tables~\ref{detections} and~\ref{nondetections}
is not a definitive statement of what is and what is not variable.

\begin{table*}
\caption[]{Variability detections. $t_{\rm max}$ is the maximum time
span of observations: the minimum span was between 10 and 20 minutes.
Two measures of variability amplitude are given: the average of the
absolute relative magnitudes, $\overline{|m_d|}$, and the
root-mean-square (RMS) of the relative magnitudes, $\sigma_m$.
$\overline{\delta m_d}$ is the average photometric error in the light
curve. $1-p$ is the probability that the variability is intrinsic to
the target. ``Obs.\ run'' gives the date (YY-MM) of the observations.
\label{detections}
}
\begin{tabular}{llrrrrrrrr}
\hline
\vspace*{0.2ex}
target		& SpT	& t$_{\rm max}$	& $\overline{|m_d|}$	& $\sigma_m$	& $\overline{\delta m_d}$	& $p$	& ~~No.\ of 	& ~~No. of 	& ~~Obs.	\\
		&	& ~~hours	& ~~mags			& ~~mags		& ~~mags				&	& frames& refs	& run	\\
\hline
2M0345		& L0	&  53	& 0.012		& 0.017		& 0.011     	&  4e-4	& 27	& 23	& 99-09  \\
2M0913  	& L3    & 125	& 0.042         & 0.055		& 0.039     	&  7e-4	& 36	& 14	& 99-01  \\
2M1145		& L1.5	& 124	& 0.026         & 0.031		& 0.022     	&  1e-3	& 31	& 12	& 99-01  \\
$''$		& $''$	&  76	& 0.015         & 0.020		& 0.012     	&~~$<$1e-9& 70	& 11	& 00-02  \\
2M1146		& L3	& 124	& 0.012         & 0.015		& 0.011     	&  3e-3	& 29	&  7	& 99-01  \\
2M1334		& L1.5	& 126	& 0.017         & 0.020		& 0.011     	&$<$1e-9& 51	& 12	& 00-02  \\
SDSS~0539	& L5	&  76	& 0.009         & 0.011		& 0.007     	&  3e-5	& 31	& 24	& 00-02  \\
SDSS~1203~~	& L3	&  52	& 0.007         & 0.009		& 0.007     	&  2e-3	& 51	& 13	& 00-02  \\
Calar 3		& M9	&  29	& 0.026         & 0.035		& 0.027     	&  6e-4	& 42	& 21	& 99-01  \\
S~Ori~31	& (M6.5)	&  50	& 0.010         & 0.012		& 0.007     	&  4e-5	& 21	& 30	& 00-02  \\
S~Ori~33	& (M6.5)	&  51	& 0.008         & 0.010		& 0.007     	&  2e-3	& 21	& 43	& 00-02  \\
S~Ori~45	& M8.5	&  50	& 0.051  	& 0.072  	& 0.032		&  5e-9 &  21	& 30	& 00-02	\\
\hline
%with last night:
%2M0030	L2	 80	0.018       0.023       0.019      6e-3     42	27 	99-09
%2M0345	L0	 77	0.015       0.016       0.010     <1e-9     31	23	99-09
%roque16M6	 77	0.012       0.014       0.011      1e-3     25	18	99-09
\end{tabular}
\end{table*}

\begin{table*}
\caption{Variability non-detections.  The columns are the same as in
Table~\ref{detections} except that here $\overline{|m_d|}$ and
$\sigma_m$ are the upper detection limits on the variability
amplitudes.  The minimum time between observations of a given target
was between 3 minutes (for 2M1439) and 35 minutes (for Roque 12).
\label{nondetections}
}
\begin{tabular}{llrrrrrrrr}
\hline
target		& SpT	& t$_{\rm max}$	& $\overline{|m_d|}$	& $\sigma_m$	& $\overline{\delta m_d}$	& $p$	& ~~No.\ of 	& ~~No. of 	& ~~Obs.	\\
		&	& ~~hours	& ~~mags			& ~~mags		& ~~mags				&	& frames& refs	& run	\\
\hline
2M0030		& L2	&  51	& 0.018		& 0.025		& 0.020		& 0.21	& 37	& 27	& 99-09 	\\
2M0326		& L3.5	&  49	& 0.021  	& 0.029		& 0.017		& 0.56	& 19	& 36	& 99-09 	\\
2M1439		& L1	&  97	& 0.007  	& 0.009		& 0.007		& 0.10	& 48	& 13	& 00-02	\\
Roque 11	& M8	& 100	& 0.028  	& 0.043		& 0.027		& 0.46	& 47	& 23	& 99-01	\\
Roque 12 	& M7.5	&  50	& 0.016  	& 0.022		& 0.015		& 0.02	& 17	& 43	& 99-09	\\
Roque 16	& M6	&  29	& 0.010  	& 0.014		& 0.010		& 0.35	& 16	& 34	& 99-09	\\
Teide 1		& M8	& 100	& 0.029  	& 0.041		& 0.030		& 0.10	& 47	& 23	& 99-01  \\
S~Ori~34~~	& (M6)	&  51	& 0.008  	& 0.010 	& 0.007		& ~~0.28	& 21	& 43	& 00-02	\\
S~Ori~44	& M6.5	&  51	& 0.030  	& 0.035		& 0.026		& 0.06	& 21	& 30	& 00-02 	\\
S~Ori~46	& (M8.5) 	&  51	& 0.032  	& 0.041 	& 0.030		& 0.03	& 21	& 43	& 00-02	\\
\hline
%with last night
%2M0326	L3.5	 77	0.020  0.025	0.017		0.06	25	34	99-09 
\end{tabular}
\end{table*}

\subsection{Comments on Individual Objects}

Notes are now given on the light curves and power spectra of objects
with statistically significant $\chi^2$ detections. Brief comments are
given at the end of the section on the non-detections.  The
implications of these results are discussed in
section~\ref{discussion}.\\

\noindent{\em 2M0345}.  The light curve shows no interesting features
and there are no peaks in the cleaned power spectrum above four times
the noise.\\

\noindent{\em 2M0913}. This detection is due primarily to a
significant drop in the flux around AJD 1187.5,
going down to 0.13 magnitudes below the median for that night. There is no
evidence for variability within the other three nights. There are no
strong periodicities in the cleaned power spectrum, the strongest
three being at 3.36, 0.76 and 0.64 ($\pm 0.08$) hours, each at around
only five times the noise level.\\

\begin{figure}
\includegraphics[angle=270,width=0.65\textwidth]{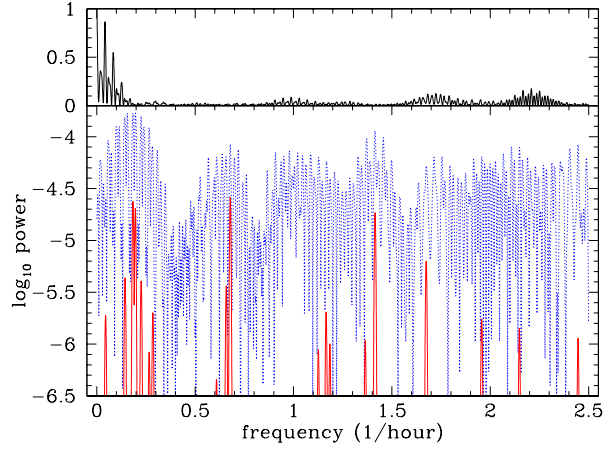}
\caption{Power spectrum for 2M1145 light curve from the 99-01 run. The
bottom panel shows the dirty spectrum (dotted line) and the cleaned
spectrum (solid line) in units of $\log_{10}(P)$. The noise level is
at about $\log_{10}(P) = -5.6$.  The top panel shows the shape of the
spectral window function on a linear vertical scale, normalised to a
peak value of 1.0.}
\label{2m1145_9901_ps}
\end{figure}

\noindent{\em 2M1145}. Evidence for variability in this L dwarf was
presented in \cite{bailer-jones99}, and it was tentatively claimed to
be periodic with a period of 7.1 hours (using the Lomb--Scargle
periodogram), pending confirmation. These data (and that for all other targets in
\cite{bailer-jones99}) have been re-reduced, and the new reduction
still shows significant evidence for variability. However, the 7.1 hour
periodicity is no longer significant in the cleaned power spectrum. 
Peaks
are present at $5.4 \pm 0.1$, $5.1 \pm 0.1$, $1.47 \pm 0.08$ and $0.71
\pm 0.08$ hours, but are only marginally significant at around eight
times the noise (Fig.~\ref{2m1145_9901_ps}). Note how difficult it
would be to confidently identify these peaks in the dirty power
spectrum.

The new reduction consists of an improved flat field, better fringe
removal, more reference stars, and a few more points in the
light curve.  The light curves from the two reductions are consistent
within their errors, but the power spectra differ, indicating that the
7.1 hour period was an artifact of higher noise and
errors which crept into the first reduction.

\begin{figure}
\includegraphics[angle=-90,width=0.65\textwidth]{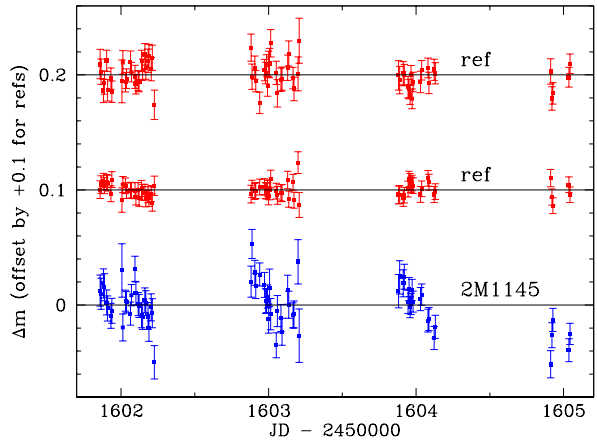}
\caption{Light curve for 2M1145 from the 00-02 run (bottom) plus a bright reference object (middle) and one of similar brightness
to the target (top).  The mean of these light
curves each offset from the mean of the target star by the amount
shown on the vertical axis. The mean for each light curve is shown as
a solid line.}
\label{2m1145_0002_lc}
\end{figure}

\begin{figure}
\includegraphics[angle=270,width=0.65\textwidth]{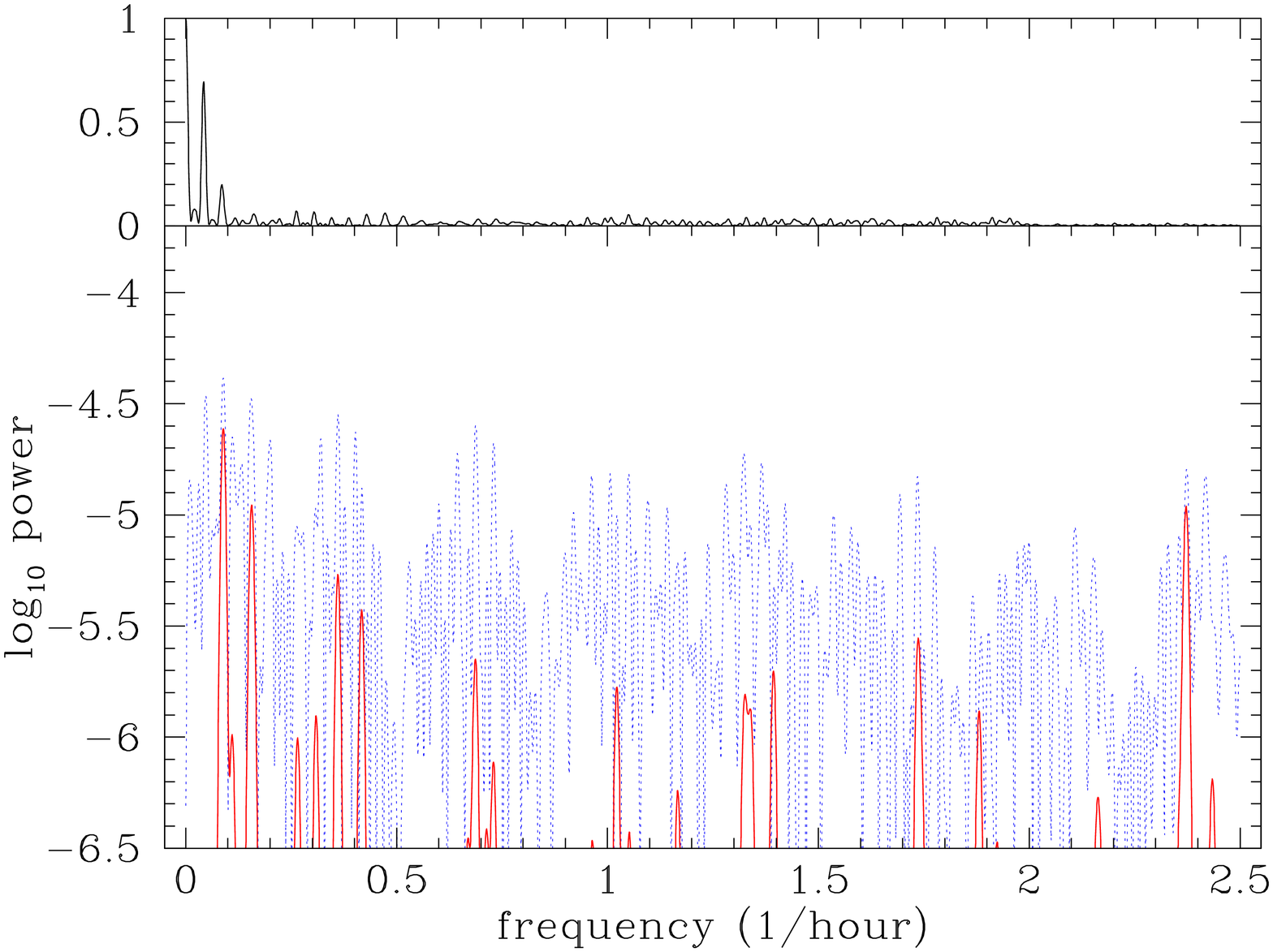}
\caption{Power spectrum for 2M1145 (from 00-02). The noise level is $\log_{10}(P) = -6.1$. See caption to Fig.~\ref{2m1145_9901_ps}.}
\label{2m1145_0002_ps}
\end{figure}
2M1145 was re-observed at higher SNR and with more epochs in the 00-02
run. These data (Fig.~\ref{2m1145_0002_lc}) also show very strong
evidence for variability, and the power spectrum shows significant
peaks at the following periods (with power in units of the noise in
parentheses): $11.2 \pm 0.8$ (31), $6.4 \pm 0.3$ (14), $2.78 \pm 0.13$
(7), $0.42 \pm 0.13$ (14) hours (Fig.~\ref{2m1145_0002_ps}).  Note
that the first period is four times the third, so these may not be
independent.  There are no common peaks in this power
spectrum and the one from 99-01.  This means that 2M1145 cannot have
both stable surface features (over a one year timescale) and a
rotation period of between 1 and 70 hours, as if it did we would have
detected it in both runs (see section~\ref{simulations}).  \\

\noindent{\em 2M1146}. The power spectrum shows peaks at the following
periods (with power in units of noise): $5.1 \pm 0.1$ (15), $3.00 \pm
0.08$ (6), $1.00 \pm 0.08$ (5), and $0.64 \pm 0.08$ (9) hours. The second and third are in the ratio 3:1, so
are probably not independent. The one at three hours is more
convincing based on the phase coverage in the phased light curve. This
is one of only two L dwarfs in the sample which has a measured $v \sin
i$, which, at 32.5$\pm$2.5\,km/s \cite{basrietal00}, implies a
rotation period of $3.7 \pm 0.3$ hours, or less,
due to the unknown inclination of the rotation axis (assuming a radius
$0.1 R_{\odot}$ \cite{chabrier00a}).  2M1146 appears to be an L
dwarf--L dwarf binary \cite{koerner99} with a separation 0.3$''$, as
well as having a background early-type star only 1''
away \cite{kirk99}. The light curve is a composite of variations in all
three objects.\\

\begin{figure}
\includegraphics[angle=-90,width=0.65\textwidth]{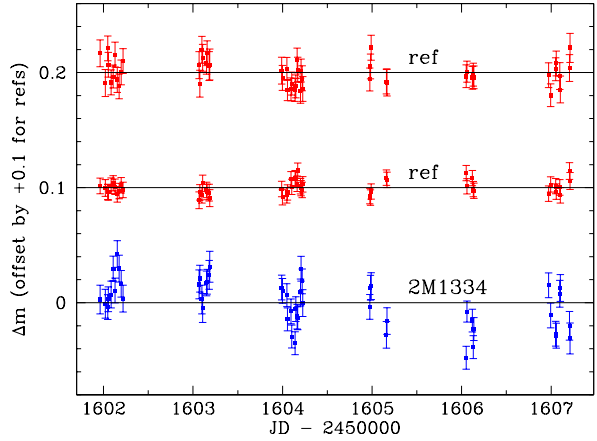}
\caption{Light curve for 2M1334 (bottom) plus a bright reference
object (middle) and one of similar brightness to the target (top). See
caption to Fig.~\ref{2m1145_0002_lc}.}
\label{2m1334_lc}
\end{figure}

\begin{figure}
\includegraphics[angle=270,width=0.65\textwidth]{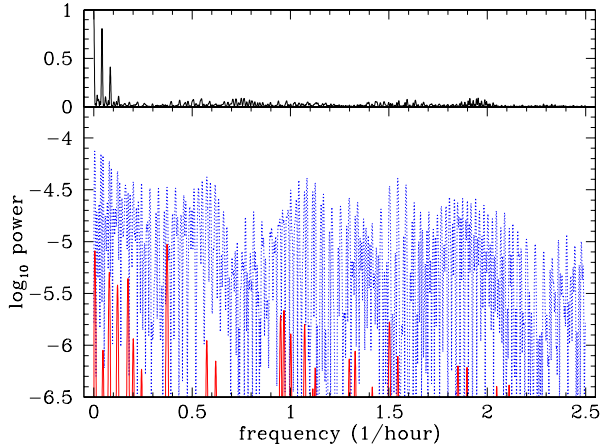}
\caption{Power spectrum for 2m1334. The noise level is $\log_{10}(P) = -6.2$. See caption to Fig.~\ref{2m1145_9901_ps}.}
\label{2m1334_ps}
\end{figure}

\noindent{\em 2M1334}. This is significantly variable, and the light
curve shows clear fluctuations within a number of nights
(Fig.~\ref{2m1334_lc}).  The largest peak in the power spectrum
(Fig.~\ref{2m1334_ps}) is at $2.68 \pm 0.13$ hours at 12 times the
noise. \\

\noindent{\em Calar 3}. The light curve shows no conspicuous features. The two
most significant peaks in the power spectrum (at 14.0 and 8.5 hours)
are less than five times the noise level, so are barely significant.\\

\begin{figure}
\includegraphics[angle=-90,width=0.65\textwidth]{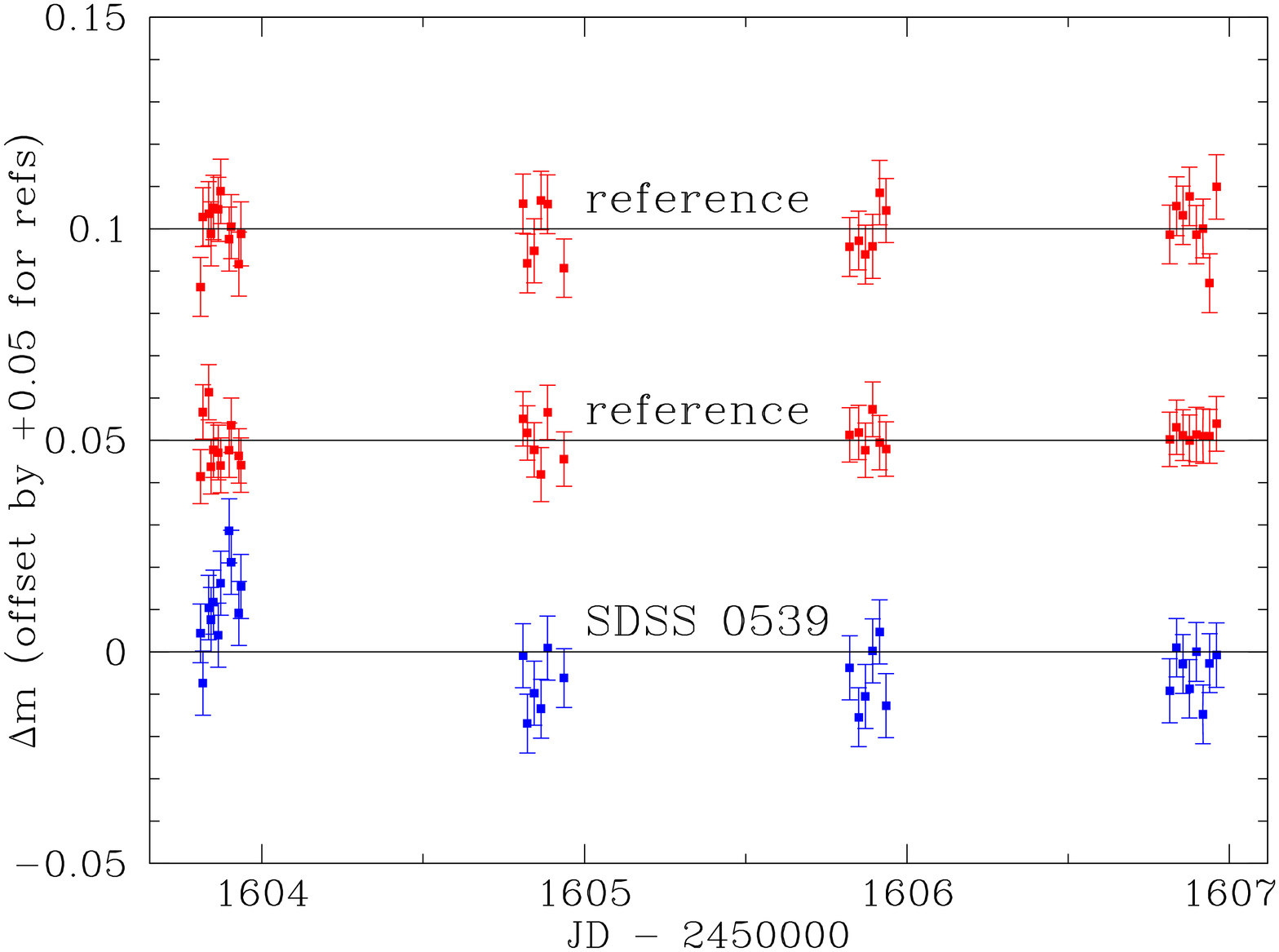}
\caption{Light curve for SDSS~0539 (bottom) plus a bright reference object (middle) and one of similar brightness
to the target (top). See caption to Fig.~\ref{2m1145_0002_lc}.}
\label{sdss0539_lc}
\end{figure}

\begin{figure}
\includegraphics[angle=270,width=0.65\textwidth]{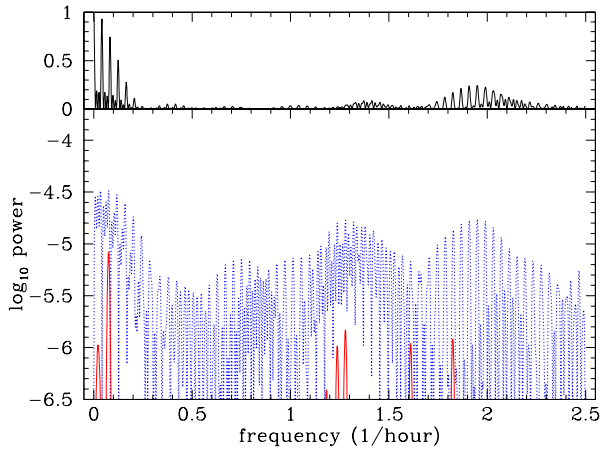}
\caption{Power spectrum for SDSS~0539. The noise level is $\log_{10}(P) = -6.4$. See caption to Fig.~\ref{2m1145_9901_ps}.}
\label{sdss0539_ps}
\end{figure}

\begin{figure}
\includegraphics[angle=-90,width=0.65\textwidth]{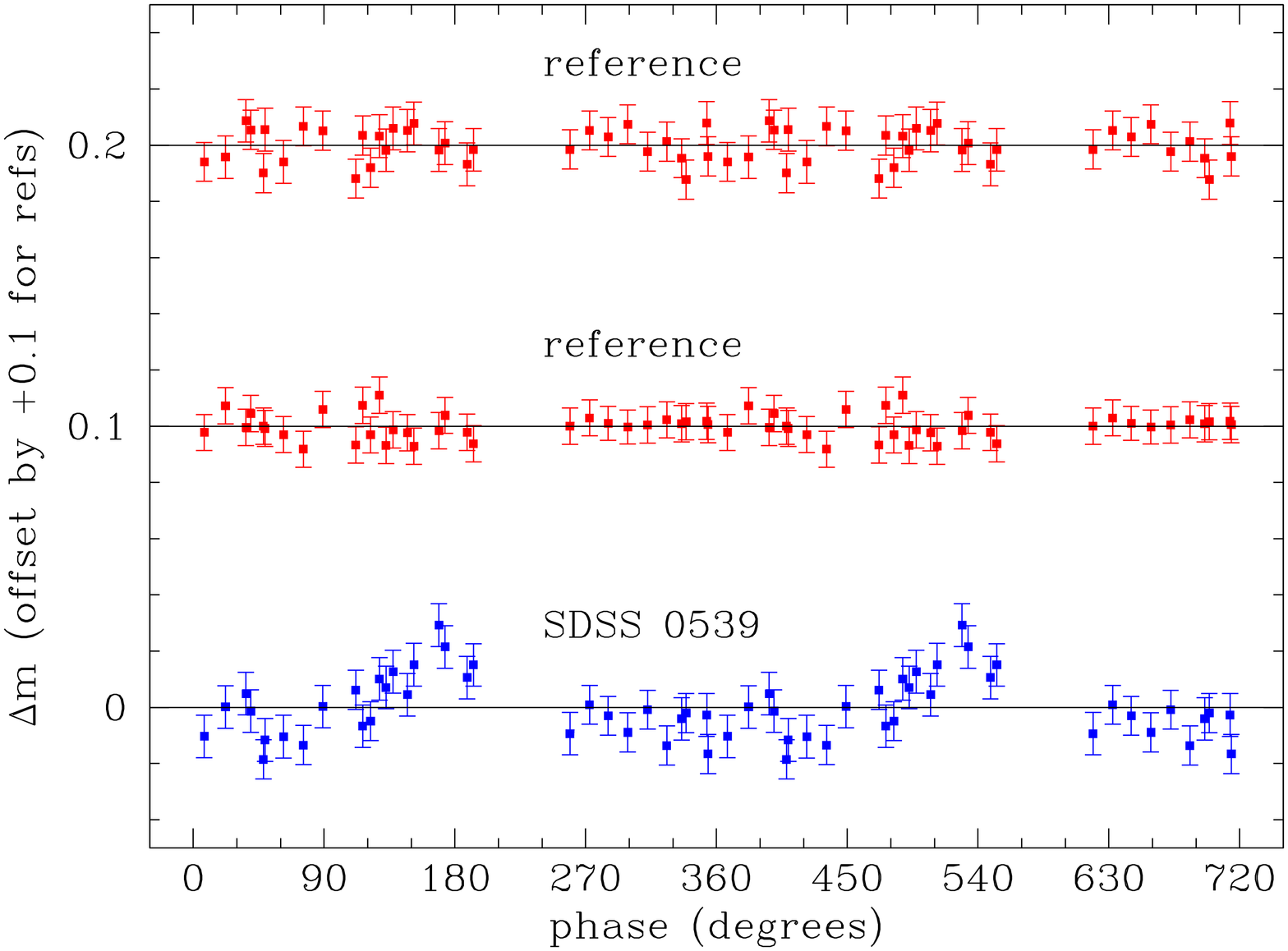}
\caption{Light curve (bottom) for SDSS~0539 phased to a period of 13.3
hours. The cycle is shown twice ($^{\circ}$--360$^{\circ}$ and
360$^{\circ}$--720$^{\circ}$).  Also
shown are two reference stars from Fig.~\ref{sdss0539_lc} phased in the same way.}
\label{sdss0539_ph}
\end{figure}

\noindent{\em SDSS~0539}. The seeing was worse than average for many
of the frames in this field, so a larger photometry aperture of radius
5.0 pixels was used. (This increases the noise and hence lowers the
sensitivity.) The significant $\chi^2$ is partly due to the brighter
points around AJD 1604. Otherwise the light curve shows no obvious
patterns (see Fig.~\ref{sdss0539_lc}). The power spectrum shows a
significant (20 times noise) peak at $13.3 \pm 1.2$ hours
(Fig.~\ref{sdss0539_ps}). The light curve phased to this period is
shown in Fig.~\ref{sdss0539_ph}. \\

\begin{figure}
\includegraphics[angle=-90,width=0.65\textwidth]{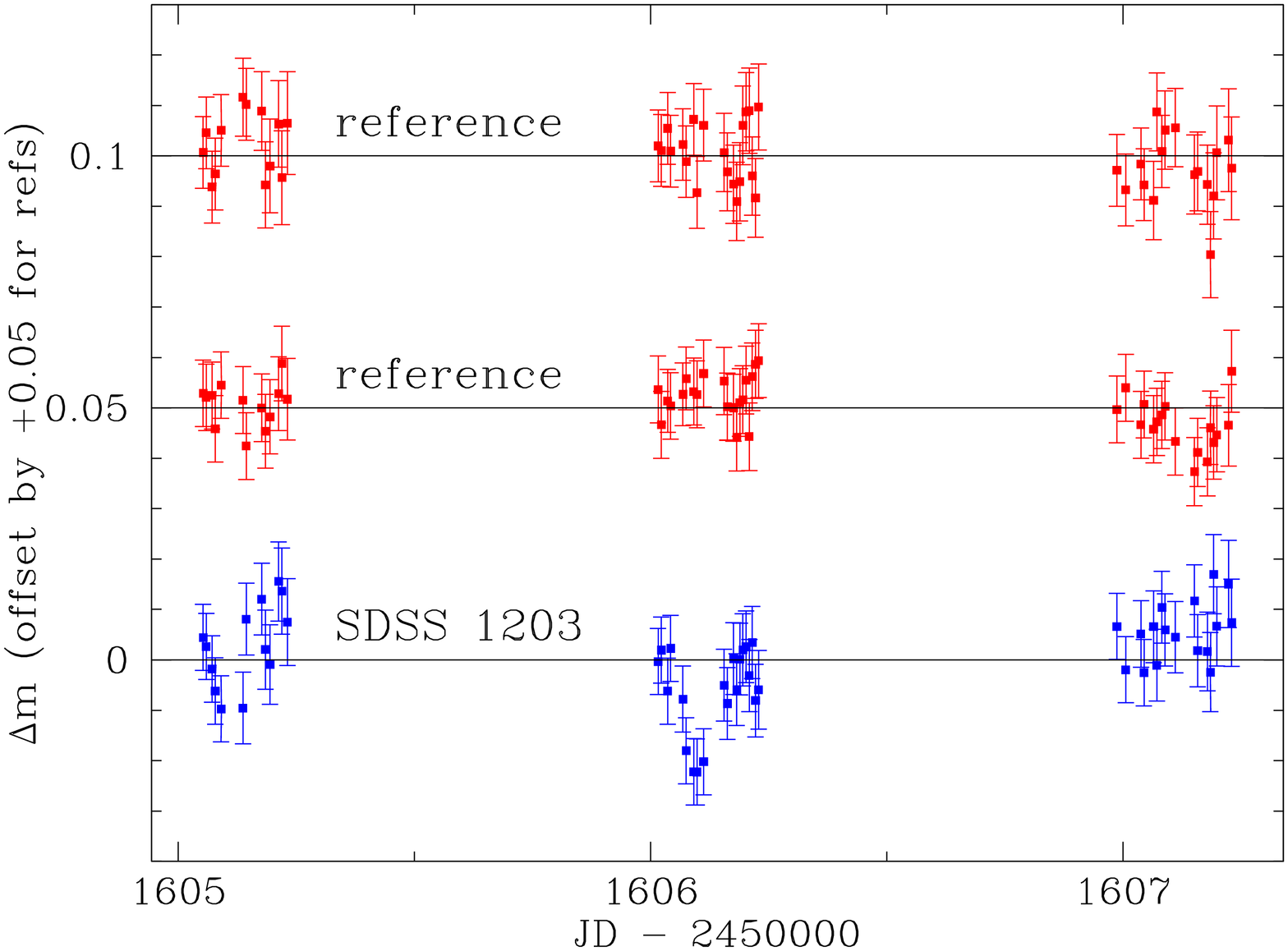}
\caption{Light curve for SDSS~1203 (bottom) plus a bright reference object (middle) and one of similar brightness
to the target (top). See caption to Fig.~\ref{2m1145_0002_lc}.}
\label{sdss1203_lc}
\end{figure}

\noindent{\em SDSS~1203}. This variability is primarily due to a drop
in brightness of about 0.02 magnitudes in four consecutive
measurements around AJD=1606.1 (Fig.~\ref{sdss1203_lc}) lasting
between one and two hours. This could be due to a short-lived surface
feature, or possibly an eclipse by a physically associated companion.
\\

\noindent{\em S~Ori~31}.  The power spectrum shows
significant peaks at $7.5 \pm 0.6$ and $1.75 \pm 0.13$ hours at 18 and
9 times the noise level respectively.  The former period dominates and
may be the rotation period.\\

\noindent{\em S~Ori~33}. The light curve shows
a rise just before AJD 1606, and the power spectrum has peaks of 6 to
 7 times the noise at $8.6 \pm 0.7$ and $6.5 \pm 0.4$ hours.  The
 former has good phase coverage, so may be the rotation period,
 although it is not a strong peak.\\

\noindent{\em S~Ori~45}.  The light curve shows three points much
lower than the average around AJD 1604.9, spanning a range of almost
0.25 magnitudes.  There is, however, a bright ($\Delta m = 1.7$)
nearby (5$''$) star which may well interfere with this variability
determination.  If these points are excluded there is no evidence for
variability ($p=0.18$).  The most significant peak in the power
spectrum is at $0.50 \pm 0.13$ hours (at 20 times the noise), which
would be extremely fast if it is the rotation period.\\

\noindent{\em Non-detections}.  2M1439 has been measured 
to have a $v \sin i$ of $10 \pm 2.5$\,km/s \cite{basrietal00},
implying a period of less than 12.1 hours for a $0.1 R_{\odot}$
radius. S~Ori~46 has a bright nearby star, which may affect the
attempt to determine variability in this object. Roque 11 and Teide 1 have also been
observed for variability in the $I$ band by Terndrup et al.\
\cite{terndrup99}.  They also did not find evidence for variability,
with measured values of $\sigma_m$ of 0.041 and 0.045 magnitudes
respectively.

\subsection{Summary of the Results}

11 of the 21 objects show evidence for variability at the
99\% confidence level ($p=0.01$).  Of these, four (2M1145, 2M1334,
SDSS~0539, S~Ori~31) show strong evidence for variability
($p<$\,1e-4). S~Ori~45 is formally a fifth object with strong evidence
for variability, but the presence of a bright close star makes us
hesitant to draw this conclusion. In four cases (2M1146, 2M1334,
SDSS~0539, S~Ori~31) we have detected dominant significant periods in
the range 3--13 hours, which may be rotation periods in all but the
first case.  S~Ori~45 also has a dominant peak, but at 0.5 hours this
would be very rapid if it is a rotation.  The remaining objects do not
show dominant periods, although the two earliest-type variables
(S~Ori~31 and S~Ori~33) show near-sinusoidal light curves at detected
periods. The light curve of one object, SDSS~1203, is essentially
featureless except for a dip which may be due to an eclipse by a
companion, although there is no direct evidence for this.

All of the objects which show variability have RMS amplitudes
($\sigma_m$ in Table~\ref{detections}) between 0.01 and 0.055
magnitudes (ignoring S~Ori~45), but most lie in the range 0.01 to 0.03
magnitudes and vary on timescales of a few hours. More detailed results are
provided in Bailer-Jones \& Mundt \cite{bjm00}.

\section{Discussion}\label{discussion}

\subsection{Simulations of the Light Curves of Rotating Spotted Stars}\label{simulations}

The power spectrum is a representation of the light curve in the
frequency domain. Specifically it gives the contributions to the
variance in the light curve of sinusoids as a function of their
frequencies. However, a significant peak in the power spectrum does
necessarily correspond to a {\it long-term} (and hence meaningful)
periodicity.  After all, {\em any} light curve -- including a random
one -- can be described in terms of its power spectrum, as all
features in the time domain must appear in the frequency domain
somehow.

In particular, the ``ideal'' case of a pure sinusoidal light curve is
only produced by a rotating star if one hemisphere is uniformly darker
than the other and the star is observed along its equatorial plane. In
contrast, a star with a single small surface feature\index{surface
features} (``spot'') would show a sinusoidal light curve only when the
spot is on the observable hemisphere; for up to half of the
rotation\index{rotation} (depending on the inclination of the rotation
axis) the light curve would be constant. A star with several spots
will show a more complex behaviour, due to the variable number of
spots observable (and hence modulating the light curve) at any one
time. All of these variations will be explained by apparent
``periods'' in the power spectrum, some of which may even be
significant relative to the noise.

\begin{figure}
\includegraphics[width=0.65\textwidth,angle=0]{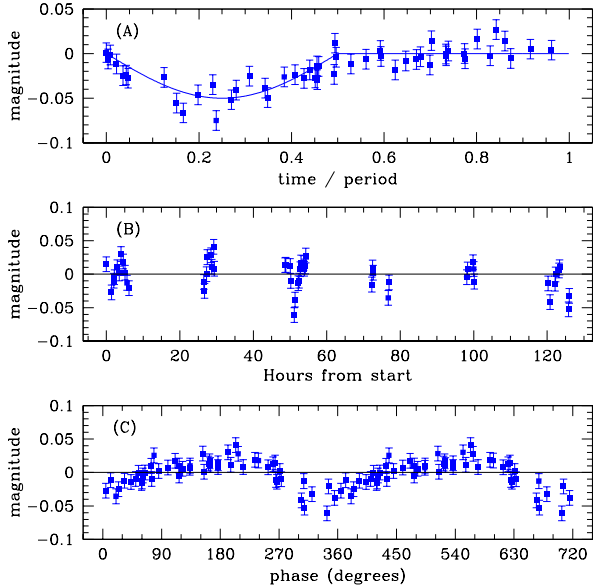}
\caption{(A) The solid line shows the true (noiseless) light curve of
a rotating star viewed equatorially with a single dark spot of
0.05 magnitudes contrast.  The star is rotated with a period of five
hours and observed as 2M1334 was (i.e.\ with the same time sampling
and with additive Gaussian noise of standard deviation 0.011
mags), giving the observed light curve in (B). (These points are
also plotted in (A) wrapped to the rotation period.)  Thus is
significantly variable according to the $\chi^2$ test ($p<$\,1e-9).
The cleaned power spectrum (Fig.~\ref{simul01_ps}) detects a period at
$5.01 \pm 0.10$ hours: the light curve phased to this detected period
{\em and phase} is shown in (C) (cycle shown twice).}
\label{simul01_rot}
\end{figure}

\begin{figure}
\includegraphics[width=0.50\textwidth,angle=270]{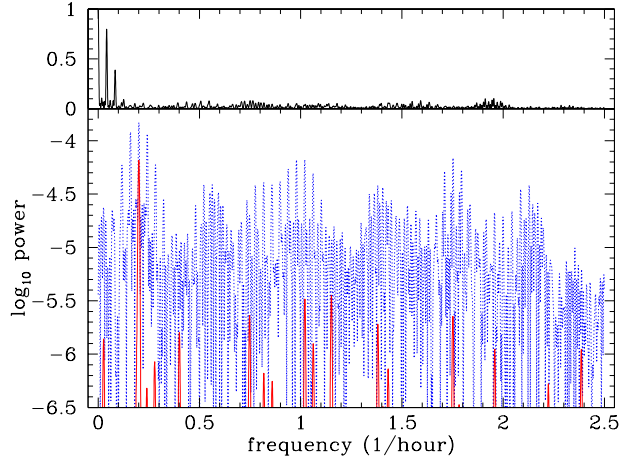}
\caption{Power spectrum for the simulated light curve shown in
Fig.~\ref{simul01_rot}B. The noise level is $\log_{10}(P) = -6.2$. The
same CLEAN parameters were used here as for the real data of
section~\ref{results}. See caption to Fig.~\ref{2m1145_9901_ps}.}
\label{simul01_ps}
\end{figure}

We have carried out numerous simulations to understand the appearance
of the light curve and its power spectrum due to such spots.
Fig.~\ref{simul01_rot} shows the light curve due to a single small dark
spot on a star.  If we rotate this star with a period of five hours
and observe it with the same noise level and time sampling as one of
the targets (2M1334), we obtain the power spectrum and phased light
curve in Fig.~\ref{simul01_ps} and Fig.~\ref{simul01_rot}C
respectively.  The phased light curve is not sinusoidal, yet the power
spectrum detects the rotation period.

\begin{figure}
\includegraphics[width=0.65\textwidth,angle=0]{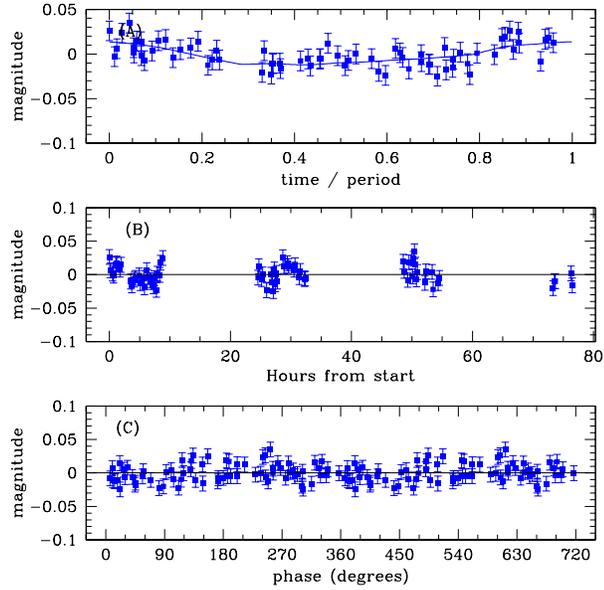}
\caption{Same as Fig.~\ref{simul01_rot} except now for eight dark and bright spots
with random phases. This gives a significant detection, although not overwhelming
($p=$0.005), yet the cleaned power spectrum (Fig.~\ref{simul07_ps})
still detects the rotation period of 10 hours.}
\label{simul07_rot}
\end{figure}

\begin{figure}
\includegraphics[width=0.65\textwidth,height=0.50\textwidth,angle=0]{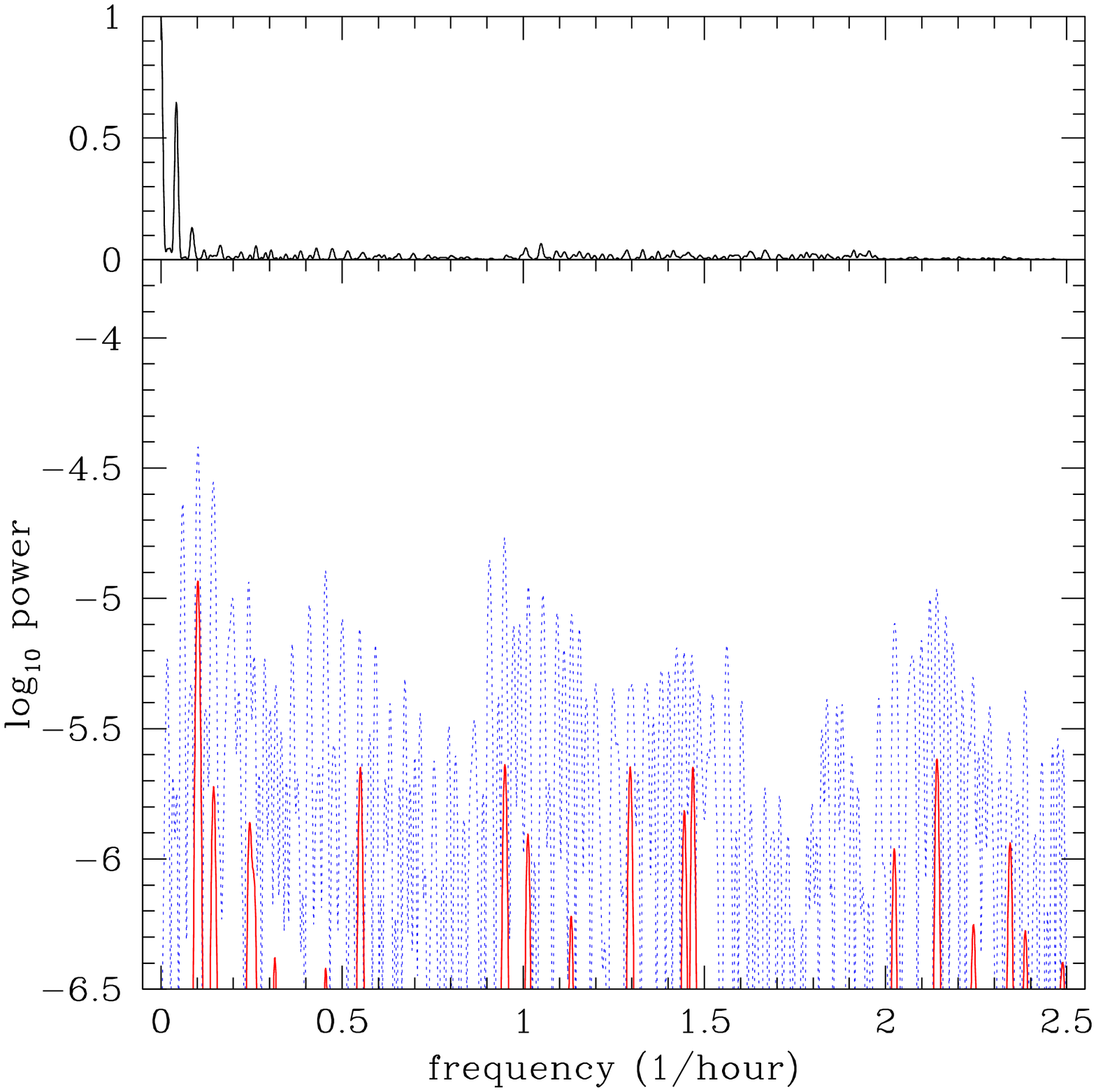}
\caption{Power spectrum for the simulated light curve shown in Fig.~\ref{simul07_rot}B. The noise level is $\log_{10}(P) = -6.1$. See caption to Fig.~\ref{2m1145_9901_ps}.}
\label{simul07_ps}
\end{figure}

Another simulation is shown in Fig.~\ref{simul07_rot}, which is due to
a star with eight spots rotating with a period of ten hours. Here the
contrast of the individual spots is much smaller, only $-0.008$ to
$+0.014$ magnitudes. The sampling and noise from 2M1145 (00-02 run) is
used and results in a significant variability detection according to
the $\chi^2$ test, but with $p$\,$=$\,0.005 is close to the
variable/non-variable cut-off. Despite this low SNR (and no detectable
sinusoidal variation in Fig.~\ref{simul07_rot}C), the rotation period
still clearly stands out in the cleaned power spectrum
(Fig.~\ref{simul07_ps}).

We see that the light curve phased to the detected rotation period
does not necessarily show sinusoidal (or even near-sinusoidal)
variation. More extensive simulations with different numbers,
amplitudes and phases of spots have been carried out.  These indicate
that even if the contrast of the spots (and hence the SNR) is very
low, then provided the light curve shows significant variation
according to the $\chi^2$ test, the rotation period is seen at more
than ten times the noise in the power spectrum.  Thus the absence of
significant periods in the power spectrum for a variable light curve
indicates {\it non-periodic} variations over the timescale of
the observations.

\subsection{Evidence for the Evolution of Surface Features}

This non-periodic variability appears to imply one
of three things:
\begin{enumerate}
\item{the rotation period is shorter than the time span of observations;}
\item{the rotation period is longer than the time span of observations;}
\item{the surface features which are presumed to be modulating the light
curve are not stable over the time span of observations.}
\end{enumerate}
The first of these implies a rotation period of less than 0.4 hours,
which corresponds to an equatorial rotation speed of at least
240\,km/s. Based on t$_{\rm max}$ in Tables~\ref{detections}
and~\ref{nondetections}, the second possibility requires {\it maximum}
$v \sin i$ values (i.e.\ when viewed equatorially) of between 1 and 4\,km/s, assuming a radius of $0.1
R_{\odot}$.  
These would be inconsistent with the measurements of Basri et al.\
\cite{basrietal00}, who report $v \sin i$ values between 10 and
60\,km/s for 16 out of 17 late M and L field dwarfs. 
(According to models, even the youngest, warmest objects
in Table~\ref{targets} -- those in $\sigma$~Orionis -- can have radii no
larger than $0.2 R_{\odot}$
\cite{chabrier00a}, so $v \sin i$ could not be above 8\,km/s for periods of order t$_{\rm max}$.)
The results of Basri et al.\ therefore imply typical rotation periods of 1 to 10
hours, and the simulations have shown that such periods would have been
detected in the light curves of the present sample, {\em if} these
objects had stable modulating surface features.  Yet some
significantly variable objects show no significant periodicities.  The
logical explanation in these cases (especially 2M0345, 2M0913, 2M1145
and Calar~3) is, therefore, that these objects have surface features
which evolve over the period of the observations, thus removing the
rotational modulation from the light curve.  For 2M1145 we possibly
have more direct evidence of this, as the two light curves from one
year apart show no common periods.

\subsection{Physical Nature of the Surface Features}

\begin{figure}
\includegraphics[width=0.65\textwidth,angle=270]{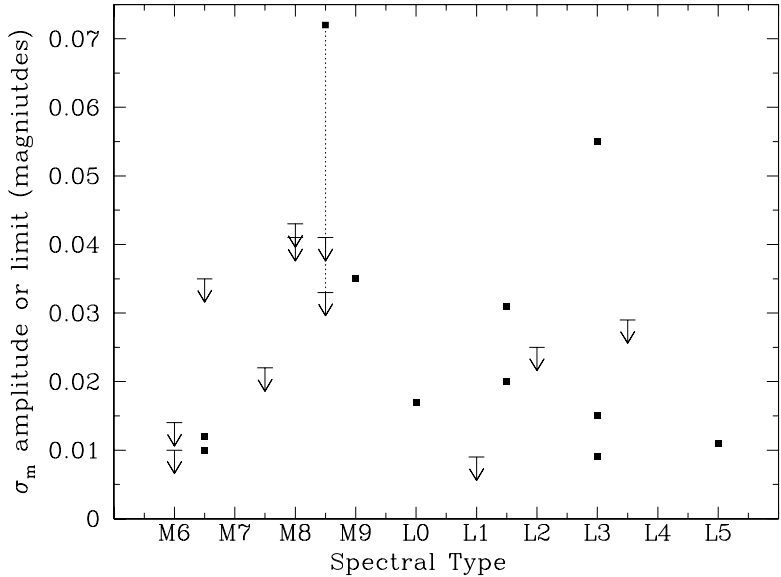}
\caption{Relationship between variability amplitudes (squares) or upper
limits to variability (arrows) and spectral type. 
S~Ori~45 (M8.5) is plotted as both an amplitude and a limit (connected with
a dotted line) depending on whether the first night of data is
included or not. The plot using $\overline{|m_d|}$ rather than $\sigma_m$
as the amplitude measure is very similar.}
\label{amps}
\end{figure}

Magnetically-induced star spots\index{magnetic spots} are common in solar-type stars, the
magnetic field being produced by the $\alpha \Omega$ dynamo. 
This appears not to operate in low mass stars and brown dwarfs,
yet a turbulent dynamo may
\cite{chabrier00a}.  However, recent observations imply that
chromospheric activity -- and, perhaps, the contrast of magnetic spots
-- decreases rapidly between spectral types M7 and L1
\cite{basri00}\cite{gizis00}.  In comparison, Fig.~\ref{amps} shows
the amplitude of variability (or upper limit thereon) as a function of
spectral type for the sample in this paper.  Whereas 7/10 of objects
later than M9 show variability, only 2/9 earlier than this do. (The
average detection limits/amplitudes are almost identical in the two
regions, so this is not an artifact.)  If the variability were due to
magnetic spots, then in the light of the activity decline we would
expect variability to be {\it less} common among later-type objects,
not {\it more} common as seen here. Although this could also be an age effect
(all of the targets of type M9 and earlier are cluster members with
ages less than 120\,Myr) it hints towards a non-magnetic origin of the
surface features.

Another candidate for producing variability is photospheric dust\index{dust}
clouds. Modelling of optical and infrared spectra show that solid
particles form in late M and L dwarfs
\cite{burrows99}\cite{chabrier00}\cite{jones97}\cite{lodders99}. While
we may expect this dust to gravitationally settle below the
photosphere, certain processes (such as tubulence) may prevent this,
and models which include dust opacity (as well as removing dust
constituents from the equation of state) give better fits to the near
infrared spectra of ultra cool dwarfs \cite{chabrier00}.  This dust could form into
large-scale opaque clouds. Their evolution over a few
rotation periods could account for the variability reported in this paper,
possibly driven by rapid rotation and turbulence.  The clouds would
have to be relatively large, as many small clouds evolving
independently would have an insignificant net effect on the light
curve. As more dust can form in cooler objects, we may expect more
variability in later-type objects, as seen in Fig.~\ref{amps}.
However, given the small amount of data on any one object, it is
premature to attempt to determine the characteristics or dynamics of
the variability patterns.

Other causes of the variability can be considered, including flaring,
accretion (for the youngest objects) or infall in an interacting
binary. However, these all rely on transient phenomena, so no one
explanation for all objects is that satisfactory.

\section{Summary}

Light curves for 21 late M and L dwarfs were obtained to probe
variability on timescales between a fraction of an hour and over 100
hours. 11 objects showed evidence for variability at the 99\%
confidence level according to a $\chi^2$ test, with amplitudes between
0.009 and 0.055 magnitudes (RMS).  The ten non-detections have upper
limits on their RMS amplitudes of between 0.009 and 0.043 magnitudes.

Power spectral analysis showed that a few objects (2M1146, 2M1334,
SDSS~0539, S~Ori~31) had significant, dominant periods between 3 and
13 hours.  For 2M1334, SDSS~0529 and S~Ori~31 these may be the
rotation periods.  The remaining seven significantly variable light
curves did not show dominant periods, and in at least three cases
(2M0345, 2M0913, Calar~3) there are not even any significant periods.
Simulations showed that any plausible period would have been detected
for these objects.  It was concluded that this non-periodic behaviour
is probably due to the evolution of surface features (assumed to
produce the variability) on timescales of a few to a few tens of
hours. These variabilities blur the rotation period, inhibiting its
detection. This is supported by observations of 2M1145 one year apart,
in which the two light curves have no common periodicities.

It was speculated that this variability (at least for the later-type
objects) is due to photospheric dust clouds, the evolution of which
could be driven by turbulence and rapid rotation. In support of this
is the greater propensity for variability in objects later than M9,
while magnetic activity (which could otherwise support
the presence of magnetically-induced spots) declines greatly beyond M7.

\section*{Acknowledgements}
The author would like to thank Harry Lehto for use of his {\sc CLEAN}
code and advice on its use. The data in this paper were
obtained with the 2.2m telescope at the German--Spanish Astronomical
Center at Calar Alto in Spain.

\end{document}